\documentclass[12pt]{article}
\begin{document}

\newcommand{\ddx}[1]{\frac{\partial}{\partial x^{#1}}} 
\newcommand{\dydx}[2]{\frac{\partial{#1}}{\partial{#2}}}
\newcommand{\dydxt}[2]{\frac{d{#1}}{d{#2}}}
\newcommand{\DD}{\Delta}
\newcommand{\GG}{\Gamma}
\newcommand{\LLL}[2]{{\Lambda^{#1}}_{#2}}
\newcommand{\tp}{\otimes}             
\newcommand{\ww}{\wedge}             
\newcommand{\la}{\langle}  
\newcommand{\ra}{\rangle}  
\newcommand{\Q}[1]{\underline{#1}} 

\newcommand{\vv}[1]{{\bf #1}}             
\newcommand{\tpp}{\tp\cdots\tp}       
\newcommand{\h}{{\mathcal H}}
\newcommand{\f}{{\mathcal F}}
\newcommand{\G}{{\mathcal G}}
\newcommand{\W}{{\mathcal W}}
\newcommand{\LL}{\Lambda}
\newcommand{\ppp}{\partial}
\newcommand{\til}[1]{\stackrel{\sim}{#1}} 

\newcommand{\be}{\begin{eqnarray}}
\newcommand{\ee}{\end{eqnarray}}
\newcommand{\bes}{\begin{eqnarray*}}
\newcommand{\ees}{\end{eqnarray*}}

\title{Variational principle for gravity in the extended phase space}
\author{Pankaj Sharan\\
Physics Department,\\ Jamia Millia Islamia,\\New Delhi 110 025, INDIA}

\maketitle

\begin{abstract}
Variational formalism in the extended phase space for fields  is applied to
gravity. It is shown that the 
requirement of invariance under arbitrary local inertial frames
implies a coupling of torsion to a 3-form of matter fields on the one hand and
to a 3-form (related to the Einstein tensor) on the other. Gravitational dynamics 
is restricted to torsion zero surface in the extended phase space for Einstein-Hilbert action.

PACS : {04.20.Fy}
\end{abstract}

The purpose of this paper is to show that torsion, even when zero, 
plays the crucial role of a mediating  field
between the stress-energy tensor of matter fields and the Einstein tensor by coupling with
both. This is done using the extended phase-space formalism recently proposed by the author.
The power and elegance of the theory of differential forms are manifest in this formalism.

\section{Extended phase space in mechanics}
The extended phase-space formalism \cite{arnold}\cite{sharan} 
can be summarized as follows. 
In mechanics, where the evolution is along the one-dimensional time, 
the dynamics is determined by the stationary Poincare-Cartan or
action 1-form $\Xi=p_idq^i-Hdt$ in the extended phase space (EPS) with coordinates $(t,q,p)$.
The physical nature of  time is that it  `never stops' and is uni-directional. That is,
for trajectories in the EPS, the time variable $t$ has non-zero derivative with respect to the 
parameter $s$ of the curve : $dt/ds\neq 0$. This condition means that the trajectories in EPS
 are {\em projectable} on the time axis. Thus EPS has a fiber bundle structure 
 with time as the base manifold and the phase-space
(with coordinates $(q,p)$) as the fiber. Evolution trajectories are {\em sections} from
the one-dimensional base manifold into the EPS.

The principle of stationary action means that if a
vector field $X$ represents the  variation of variables in the EPS, then
the Lie derivative 
of action integrated over a proposed trajectory is zero : $L_X\int\Xi=0$.
Using $L_X=i_X\circ d + d\circ i_X$
we see that we should first calculate $d\,\Xi$
and obtain equations of motion by contracting (i.e. applying $i_X$) with different 
independent variation fields $X$.  The second term simply gives the Noether theorem 
$d(i_X\,\Xi)=0$ for symmetry fields $X$ which satisfy $L_X\Xi=0$.  
We can write
\bes -d\,\Xi=dq^i\ww dp_i+dH\ww dt=\left(dq^i-\dydx{H}{p_i}dt\right)
\ww\left(dp_i+\dydx{H}{q^i}dt\right) \ees 
Contracting with independent variations fields 
$X= \ppp/\ppp q^i, \ppp/\ppp p_i, \ppp/\ppp t$ etc gives the
Hamilton  equations.

\section{Co-frames and frame-gauge invariance}
Einstein's theory of gravitation regards spacetime as a four dimensional continuum
with a metric  tensor whose components $g_{\mu\nu}, \mu,\nu=0,1,2,3$ determine all
 gravitational phenomena at the classical level.

In place of the ten metric components we may choose at each spacetime point
a local inertial frame or co-frame as follows. We choose, in a smooth way,  a $4\times 4$ 
non-singular real  matrix $N$  with elements 
${N^a}_\mu,a,\mu=0,1,2,3$ at each point.
Let $\eta_{ab}$ be the matrix $\|\eta\| $ with $(-1,1,1,1)$ on the diagonal and zero elsewhere. Now define
\be g_{\mu\nu}=\eta_{ab}{N^a}_\nu{N^b}_\mu,\qquad\mbox{or}\qquad  \|g\|= N^T\|\eta\| N. \label{defg} \ee
As  defined $g_{\mu\nu}$ are elements of  a symmetric non-singular matrix 
$\|g\|$ with ($\det \|g\|=-(\det N)^2 \neq 0$)
 therefore they are qualified to represent  the components of a  metric.
However, there are many matrices $N$'s giving the same $g_{\mu\nu}$.
If we replace ${N^a}_\nu$ by $\LLL{a}{b}{N^b}_\nu$ where $\LL$ is a Lorentz transformation
$\eta_{ab}\LLL{a}{c}\LLL{b}{d}=\eta_{cd}$, the matrix $g_{\mu\nu}$ does not change. Thus there is
a six-parameter  Lorentz group `degeneracy' in choosing the matrix $N$ at each point. 
The 1-forms $n^a\equiv {N^a}_\mu dx^\mu$ 
constitute an orthonormal co-frame 
with respect to the metric $g_{\mu\nu}$  by design,  
because if we write
the equation (\ref{defg}) as $N\|g\|^{-1}N^T= \|\eta\|^{-1}$, and denote the elements of the inverse 
matrices by $g^{\mu\nu}$ and $\eta^{ab}$ then
\be \la n^a, n^b\ra\equiv   g^{\mu\nu}{N^a}_\mu {N^a}_\nu =\eta^{ab}. \ee 
Thus we  can describe the gravitational field by giving a system of orthonormal co-frames. The price to pay is  a six parameter gauge invariance.

The essence of general theory of relativity is that physics is independent of the choice of 
local inertial frames represented here by orthonormal co-frames.  We call this invariance  the ``frame gauge invariance".

A local gauge group, in turn, involves its own connection forms or gauge potentials. 
Since all physical quantities are
defined in terms of the local inertial frames, the rate of  change of these quantities
involves derivatives $dn^a$. These exterior derivatives are not frame-gauge invariant.
They have to be replaced by covariant derivatives
$dn^a+{\omega^a}_b\ww n^b$ 
where the connection matrix ${\omega^a}_b$ of 1-forms, under the frame change by Lorentz
transformation $\LL$ : $n'=\LL n$, transforms as $\omega'=\LL\omega\LL^{-1}+\LL d\LL^{-1}$.
(We omit indices and the wedge symbol when there is no confusion.)
The covariant derivatives of the co-frame $n^a$ are called the {\em torsion} 2-forms 
of the Einstein-Cartan theory :
\be T^a=dn^a+{\omega^a}_b\ww n^b \label{tor}.\ee
 The frame gauge fields, or {\em curvature 2-forms}, are determined by
\be {\Omega^a}_b\equiv d{\omega^a}_b+{\omega^a}_c\ww{\omega^c}_b. \label{curv} \ee
The torsion and curvature ((\ref{tor}) and (\ref{curv})) forms yield the {\em Bianchi identities} on exterior differentiation,
\be dT +\omega T &=& \Omega n, \label{btor}\\
d\Omega +\omega\Omega&=& \Omega\omega. \label{bcurv}\ee

The nature of the gauge group (Lorentz group here) is reflected in the antisymmetry
of the connection when one index is lowered by Minkowski metric $\eta$,
\be \omega_{ab}\equiv \eta_{ac}{\omega^c}_b=-\omega_{ba},\qquad
 \Omega_{ab}\equiv \eta_{ac}{\Omega^c}_b=-\Omega_{ba}. \ee
We will use constant matrices $\eta$'s for raising and lowering of indices.

\section{EPS for fields}

It has been suggested by the author that the fiber bundle 
structure of the extended phase space should be taken over for field
systems by replacing the 1-dimensional base manifold 
by the 4-dimensional space-time.
The Poincare-Cartan form (or action) is now a 4-form \cite{sharan}.
Let $\phi$ be a scalar field. Then
 its canonical momentum  is a differential 
1-form $p$ in this formalism. The Poincare-Cartan form has the structure
\be \Xi_\phi=(*p)\ww d\phi -H \ee
where the Hodge star operator 
is used to convert a 1-form $p$ into the 3-form $*p$
in order to obtain a 4-form of the type `$pdq$'. (See \cite{book} for definition and notational convention.) 
Here, the `covariant Hamiltonian' $H=H(\phi,p)$, is {\em not} the 3-form of energy 
density, but a 4-form
constructed from coordinate  (0-form $\phi$) and its canonical momentum (1-form $p$).
There is no `velocity' ($d\phi$) in $H$, but, in principle,  
there could be a linear term in velocity;
as is actually the case in gravity. 
 The standard scalar field Hamiltonian is,  
\be H=\frac{1}{2}(*p)\ww p+\frac{1}{2}m^2\phi^2(*1).\ee
Here we use the notation $*1=n^0\ww n^1\ww n^2\ww n^3$ for the Hodge dual
of the constant function $1$.

The variational principle in the extended phase-space for fields requires that allowed
field configurations are those four-dimensional sub-manifolds
of the extended phase space which are sections (i.e. fiber respecting mappings from 
the base into the  EPS) and on which the 4-form $i_X\circ d\,\Xi_\phi=0$
for variational field $X$.
We can write
\bes -d\,\Xi_\phi &=& 
-(d*p)\ww d\phi+(d*p)\ww p +m^2\phi d\phi\ww(*1)+\frac{1}{2}m^2\phi^2d(*1). \ees
Using the formulas
\bes p=p_an^a,\ \  dp=dp_an^a+p_adn^a,\ \  d(*p)=dp_a\ww (*n^a)+p_ad(*n^a) \ees
\bes d(*n^a)=*(n^a\ww n^b)\ww dn_b, \ \ 
(*n^a)\ww n^b=-\eta^{ab}(*1), \ \ 
d(*1)=(*n^a)\ww dn_a \ees
we calculate
\be -d\Xi_\phi = \big[dp_a(*n^a)-m^2\phi(*1)\big]\big[p-d\phi\big]-{\Theta}^b T_b \label{matter}\ee
where we replace $dn_b$ in the last term by the torsion 2-form 
$T_a=dn_a+\omega_{ab}n^b$
because the added term containing 5 factors of 1-forms $n$ in a four dimensional space is zero. 
The 
{\em stress-energy 3-form} $\Theta$ is given by
\be {\Theta}^b &=& p_a(*n^an^b)d\phi+\frac{1}{2}p_ap^a(*n^b)
-\frac{1}{2}m^2\phi^2(*n^b) \nonumber \\
&=&\left[p^bp^c-\eta^{bc}\left(\frac{1}{2}p_ap^a+\frac{1}{2}m^2\phi^2\right)\right](*n_c)\\
&\equiv& T^{bc}(*n_c) \label{stress}\ee
where $T^{ab}$ is the familiar {\em stress-energy tensor}.  The Hamiltonian equations
for matter fields from the variation of fields $\phi$ and $p$ while keeping frame fields and connection fixed are read off from the two factors  of the first term :
\be p=d\phi,\qquad dp_a(*n^a)-m^2\phi(*1)=0. \ee
If we substitute $p=d\phi$ in the second equation we obtain the Klein Gordon 
equation for $\phi$ in the curved background determined by the 
fixed frame field $n^a$. 

It is worth emphasizing that the exterior derivative acts on the full phase-space,
and $d\phi$ is independent of $n^a$ (or $dx^\mu$). The momentum is a differential 1-form
$p=p_an^a$ but $p_a$ are not functions of $x$. They are coordinates in the extended phase-space just as in classical mechanics $p_i$ are coordinates in the cotangent bundle $T^*(Q)$.
Only when we look for a section
which makes action stationary, will independent phase-space coordinates $p_a$ 
become functions of $x$. 

The term $-\Theta^aT_a$ containing torsion of the gravitational field will be seen below
to combine with the gravitational part of action when the frame field and connection
are varied.

\section{Gravitational field in EPS}

The gravitational field is jointly determined by the co-frame fields $n^a$ and the  frame-gauge connection fields $\omega_{ab}$. The ``velocities" for these are
the torsion $T=dn+\omega n$ and curvature  $\Omega=d\omega +\omega\omega$ respectively.
Since action is gauge invariant, the true variation fields $X$ of co-frames $n$ and 
gauge potentials $\omega$ contract with gauge-covariant $T$ and $\Omega$ and not $dn$ or 
$d\omega$.
  
We take as the Poincare-Cartan 4-form for gravity the Einsten-Hilbert action,
\be \Xi_{EH}= - \frac{1}{2\kappa} *(n^an^b)\Omega_{ba}=\frac{1}{4\kappa} n_aE^a,\qquad 
\kappa=8\pi G_N\ee
where $G_N$ is Newton's gravitational constant and 
where we define the {\em Einstein 3-form} as
\be E^a=*(n^a\ww n^b\ww n^c)\Omega_{bc} \equiv -2G^{ab}(*n_b). \label{EG}\ee
The components $G^{ab}$ are defined by this equation. They reduce to
 the usual symmetric Einstein tensor for the Riemannian geometry.

The variation of Einstein-Hilbert action is simply
\be d\,\Xi_{EH}=\frac{1}{2\kappa}\,E\,T.  \ee
The proof goes as follows : Using formulas such as  
$d*(n^an^b)=*(n^an^bn^c)dn_c$ and the Bianchi identity,
\bes d[*(n^an^b)\Omega_{ba}] &=& *(n^an^bn^c)dn_c\Omega_{ba} + *(n^an^b)d\Omega_{ba}\\
&=&  *(n^an^bn^c)\Omega_{ba}(T_c-\omega_{cd}n^d) + 
*(n^an^b)\big(\Omega\omega-\omega\Omega\big)_{ba}.\ees
Furthermore, as
\bes 
*(n^an^bn^c)n^d=-\eta^{ad}*(n^bn^c)+\eta^{bd}*(n^an^c)-\eta^{cd}*(n^an^b),
\ees
we can rearrange and simplify the terms not containing $T$. They all add up to 
\bes *(n^an^b)(\Omega\omega+\omega\Omega)_{ba}=
*(n^an^b)(\Omega_{bd}{\omega^d}_a+\Omega_{ad}{\omega^d}_b)=0,\ees
being a sum over a product of a symmetric and an antisymmetric expression.

Combining with the derivative of the matter action (\ref{matter})  for the scalar field
\be -d(\Xi_\phi+\Xi_{EH}) =
\big[dp(*n)-m^2\phi(*1)\big]\big[p-d\phi\big]  -\big[E/2\kappa+\Theta\big]T  
\ee
The allowed field configurations given by the terms above can be read off easily. For variational
 fields in the direction of $\phi$ and $p$ we get the Klein-Gordon equations in curved 
 background (determined by $n$ and $\omega$). In the second term varying $n$ but keeping
 $\omega$ constant gives the Einstein equation
\be  E = -2\kappa\Theta \qquad \mbox{or}\qquad G^{ab}=(8\pi G_N) T^{ab},\ee
using (\ref{EG}) and (\ref{stress}).
And lastly, if $\omega$ is varied but $n$ is kept fixed, we get the equation $T=0$.


\end{document}